\documentclass[journal]{IEEEtran}
\ifCLASSINFOpdf

\usepackage{setspace, rotating,url}
\usepackage{amsmath,amssymb,amsfonts}
\usepackage{algorithmic}
\usepackage{graphicx}
\usepackage{textcomp}
\usepackage{caption}
\usepackage{subfigure}
\usepackage{multirow}
\usepackage{array}
\usepackage{epstopdf}
\usepackage{tabularx}
\usepackage{booktabs}
\usepackage{colortbl,color}

\usepackage{cite}
\usepackage{hyperref}
\begin{document}
%
\title{ESND: An Embedding-based Framework for Signed Network Dismantling}

\author{Chenwei Xie,
Chuang Liu,
        Cong Li, 
        Xiu-Xiu Zhan, 
        Xiang Li
        
\thanks{Chenwei Xie and Chuang Liu are with the Research Center for Complexity Sciences, Hangzhou Normal University, Hangzhou 311121, China (e-mail: xiechenwei@stu.hznu.edu.cn; liuchuang\_1985@126.com).}

\thanks{Cong Li is with the Adaptive Networks and Control Laboratory, Electronic Engineering Department, School of Information Science and Engineering,  and the Research Center of Smart Networks and Systems, Fudan University, Shanghai 200433, China (e-mail: cong\_li@fudan.edu.cn, Corresponding author). }

\thanks{Xiu-Xiu Zhan is with the Research Center for Complexity Sciences, Hangzhou Normal University, Hangzhou 311121, China, and the College of Media and International Culture, Zhejiang University, Hangzhou 310058, China (e-mail: zhanxiuxiu@hznu.edu.cn, Corresponding author).}

\thanks{Xiang Li is with the Institute of Complex Networks and Intelligent Systems, Shanghai Research Institute for Intelligent Autonomous Systems, the Frontiers Science Center for Intelligent Autonomous Systems, and the State Key Laboratory of Intelligent Autonomous Systems, Tongji University, Shanghai 201210, China (e-mail: lix2021@tongji.edu.cn).}
\thanks{Manuscript received xxx; revised xxx}}
\maketitle

\begin{abstract}

Network dismantling aims to maximize the disintegration of a network by removing a specific set of nodes or edges and is applied to various tasks in various domains, such as cracking down on crime organizations, delaying the propagation of rumors, and blocking the transmission of viruses. Most of the current network dismantling methods are tailored for unsigned networks, which only consider the connection between nodes without evaluating the nature of the relationships, such as friendship/hostility, enhancing/repressing, and trust/distrust. We here propose an embedding-based algorithm, namely ESND, to solve the signed network dismantling problem. The algorithm generally iterates the following four steps, i.e., giant component detection, network embedding, node clustering, and removal node selection. To illustrate the efficacy and stability of ESND, we conduct extensive experiments on six signed network datasets as well as null models, and compare the performance of our method with baselines. Experimental results consistently show that the proposed ESND is superior to the baselines and displays stable performance with the change in the network structure. Additionally, we examine the impact of sign proportions on network robustness via ESND, observing that networks with a high ratio of negative edges are generally easier to dismantle than networks with high positive edges.
\end{abstract}


%
\IEEEpeerreviewmaketitle

\section{Introduction}
\IEEEPARstart{N}{etwork} dismantling aims to remove a certain number of nodes that could maximize the damage to the network in terms of connectivity ~\cite{albert2000error,holme2002attack,braunstein2016network}. It has become a prominent topic in network science due to its extensive applications in different fields~\cite{cohen2003efficient, albert2004structural, verma2014revealing}. For instance, it could be used to delay the spread of diseases by immunizing (or isolating) the critical nodes in epidemic-spreading networks~\cite{li2012bounds,zhan2022exploring,akhtar2023nrand}. In terms of information dissemination, it has the potential to help block key users to control the propagation of rumors and false information on online social platforms~\cite{zhan2020suppressing, gao2024efficient}. In addition, effective network dismantling measures can achieve the purpose of quickly thwarting the crime for terrorist organization networks~\cite{duijn2014relative,collins2022new}. 

Network dismantling has been proven to fall into the category of NP-hard problems~\cite{bui1992finding,buldyrev2010catastrophic,osat2017optimal}, the mathematical essence of which is a combinatorial optimization problem. Researchers have proposed various methods to identify critical nodes for network dismantling problems, such as centrality-based methods (e.g., degree, k-shell, betweenness, and closeness)~\cite{sun2016impact,zhou2021robustness,almeira2021explosive,hao2020cascading,iyer2013attack}, heuristic algorithms (e.g., acquaintance immunization, collective influence (CI) and generalized network dismantling (GND))~\cite{zhao2018optimal,morone2015influence, ren2019generalized,feng2023generalized}, meta-heuristic algorithms (e.g., artificial bee colony algorithm, memetic algorithm)~\cite{lozano2017optimizing, wang2019finding}, and machine learning algorithms (e.g., finding key players in networks through deep reinforcement learning (FINDER), graph dismantling with machine learning (GDM), neural extraction framework for multiscale essential structures (NEES))~\cite{fan2020finding,grassia2021machine, liu2022neural}. Although these methods have shown efficacy in rapidly disintegrating networks, most of them are tailored to unsigned networks, i.e., networks without positive or negative signs on the edges. Actually, interactions between different individuals in the real world may contain specific meanings~\cite{leskovec2010signed,yan2022hypernetwork,qi2023robustness}. For example, users could be friends or enemies in social networks, and a signed network is needed to represent the different relationships between users~\cite{tang2016survey}. Moreover, the dynamics of signed networks is quite different from that of unsigned networks. For instance, we need to consider signs when modeling the spread process on a signed network, and the signed network structure may result in different dynamic behaviors~\cite{li2021dynamics}. With regard to the dismantling problem, few works have considered this problem on signed networks, and the main challenge relies on how to utilize the signed network topology to solve this problem.

To address the challenge of signed network dismantling, we propose an algorithm named the Embedding-based framework for signed network dismantling (ESND), which integrates node embedding~\cite{osat2023embedding} and node clustering to achieve rapid disintegration of a signed network. The ESND consists of three main parts that iteratively remove nodes from the network (see Figure~\ref{fig1}): First, we perform a signed network embedding algorithm (SiNE) to obtain node embedding vectors that could capture the local and global structure of a signed network. Second, we employ the K-means algorithm to classify the nodes into different clusters. Lastly, the node with the highest degree in the largest cluster is removed from the network. We compare ESND with the baselines on different empirical signed networks and their null models. The results show that ESND could better dismantle a signed network than the baseline methods. 

 The subsequent sections of this paper are organized as follows. Section \ref{2} details the specifics of our proposed algorithm. Section \ref{3} offers a clear description of the baseline methods. Section \ref{4} introduces the datasets and presents all the experimental results. We summarize our work and discuss future research directions in Section \ref{5}.

\section{Methods}
\label{2}
\begin{figure*}[!htp]
    \centering
        \includegraphics[width=1.0\linewidth]{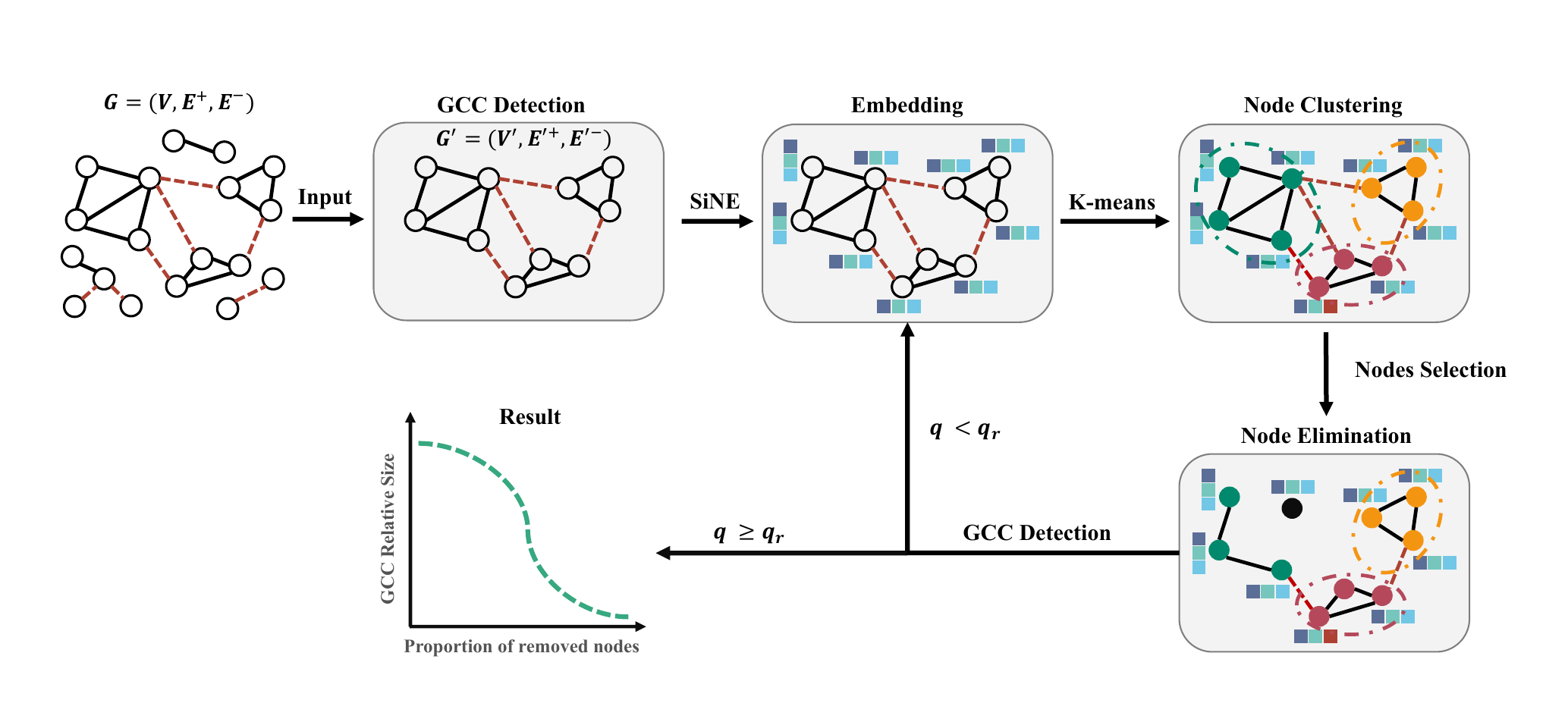}
    \caption{Framework of ESND. The solid black lines represent positive edges, the red dashed lines indicate negative edges, $q$ represents the fraction of the removed nodes, and $q_r$ is a threshold value indicating when we will stop the algorithm.}
    \label{fig1}
\end{figure*}

In this section, we introduce the iterative dynamic approach to the dismantling of signed networks, as shown in Figure~\ref{fig1}.  Initially, we identify the giant connected component (GCC) in the network and then use a signed network embedding algorithm, i.e., SiNE~\cite{wang2017signed},  to get the embedding vector of each node. Later, we use the K-means algorithm to partition the GCC into several clusters based on the embedding vectors of the nodes. The node with the highest degree in the largest cluster is removed from the network. If the network contains several nodes with the same value of the highest degree, we randomly choose one of them to remove. Subsequently, we re-identify the GCC within the remaining network and perform signed network embedding on the GCC. We then eliminate the node with the highest degree in the largest cluster using K-means. The process will iterate until the fraction $q$ of removed nodes reaches a specified value $q_r$. The essential steps, i.e., giant component detection, signed network embedding (SiNE), node clustering, and node elimination, of the ESND algorithm are illustrated as follows.

\subsection{Giant Connected Component Detection}
Given an undirected and unweighted signed network $G=(V, E)$ consisting of $N$ nodes and $M$ edges, where $V=\left \{ v_{1},v_{2},\cdots,v_{N}\right \}$ represents the set of nodes and $E$ is the set of edges. An edge $e_{ij}=(v_i, v_j) \in E$ can take a value of $1$ or $-1$, indicating a positive or negative edge in the network. To effectively dismantle a signed network, we need to detect the GCC from the current network as input for embedding at each iteration. Therefore, we use the breadth-first-search (BFS) algorithm to detect the giant connected component within a signed network. Specifically, we start from each unvisited node to find all nodes connected to it and record the size of its corresponding connected component. The component containing most nodes is referred to as the GCC.

\subsection{Signed Network Embedding (SiNE)}
We choose to use a classic signed network embedding method rooted in deep learning, specifically known as SiNE, to obtain embedding vectors for each node. In the subsequent sections, we provide an in-depth description of the three fundamental components of this method, i.e., the establishment of the objective function, the construction of a deep learning network, and the update of the parameters. 
The formulation of the objective function in SiNE is based on structural balance theory, positing that individuals are more like their ``friends'' than their ``enemies''. We utilize $\mathcal{T}=\{(v_i,v_j,v_k)\mid e_{ij}=1,e_{ik}=-1,v_i,v_j,v_k\in V\}$ to denote a collection of triplets, where there is a positive connection between $v_i$ and $v_j$, and a negative connection between $v_i$ and $v_k$. Hence, it is necessary to allocate a greater similarity to $v_i$ and $v_j$ compared to $v_i$ and $v_k$. Mathematically, we express the similarity as $f(\mathbf{x}_{i},\mathbf{x}_{j})\geq f(\mathbf{x}_{i},\mathbf{x}_{k})+\epsilon$, where $f$ denotes the similarity function that requires learning, and $\epsilon$ fine-tunes the dissimilarity between the nodes. The higher value of $\epsilon$ makes $v_i$ and $v_j$ closer and $v_i$ and $v_k$ farther away in the embedding space. Since the mentioned function is unable to handle cases where 2-hop networks of nodes only have positive or negative links, and given that positive connections are more prevalent than negative ones in real-world networks, the study introduces a virtual node $v_0$. The virtual node is utilized to establish a negative link between $v_0$ and the node connected to its 2-hop neighbors only by positive links. Assuming $\mathcal{T}_0=\{(v_i,v_j,v_0)\mid e_{ij}=1,e_{i0}=-1\}$ is one of these triplets, we have $f(\mathbf{x}_{i},\mathbf{x}_{j})\geq f(\mathbf{x}_{i},\mathbf{x}_{0})+\epsilon_0$, where $\epsilon_0$ plays a similar role as $\epsilon$. Consequently, the objective function for signed network embedding is


\begin{equation}\small
\begin{aligned}
\min_{\mathbf{X}, \mathbf{x}_0, \epsilon}
& \frac{1}{T} \left[
\sum\limits_{\left(\mathbf{x}_i, \mathbf{x}_j, \mathbf{x}_k\right) \in \mathcal{T}} \max \left(0, f\left(\mathbf{x}_i, \mathbf{x}_k\right)+\epsilon-f\left(\mathbf{x}_i, \mathbf{x}_j\right)\right) \right. \\ 
& \left. + \sum\limits_{\left(\mathbf{x}_i, \mathbf{x}_j, \mathbf{x}_0\right) \in \mathcal{T}_0} \max \left(0, f\left(\mathbf{x}_i, \mathbf{x}_0\right)+\epsilon_0-f\left(\mathbf{x}_i, \mathbf{x}_j\right)\right) \right] \\[8pt]
& + \lambda\left({H}(\phi)+\|\mathbf{X}\|_F^2+\left\|\mathbf{x}_0\right\|_2^2\right),
\end{aligned}
\end{equation}

where the size of the training data is denoted by ${T}= \left| \mathcal{T} \right|+ \left| \mathcal{T}_0 \right|$, and $\mathbf{X}=\left\{\mathbf{x}_1, \mathbf{x}_2, \cdots, \mathbf{x}_N\right\}$ represents the embedding vectors of the $N$ nodes. The similarity function $f$ is determined by the parameter set $\phi$, and $H(\phi)$ serves as a regularizer to prevent overfitting. The parameter $\lambda$ is utilized to control the impact of the regularizers.
In addition, $\|\cdot\|_F$ is the Frobenius norm, while $\|\cdot\|_2$ represents the $\ell_2$-norm.

The optimization of the objective function is carried out to acquire nonlinear embedding vectors for nodes within signed networks. Within the SiNE framework, the function $f$ and the parameter set $\phi$ in the objective function are defined through the construction of a neural network. The framework consists of two layers of neural networks, where $\mathbf{W}^{11}$ and $\mathbf{W}^{12}$ are the weights of the first hidden layer, and $\mathbf{b}^1$ is the bias. The specific output form of the first layer is as follows:
\begin{equation}
\begin{aligned}
\mathbf{z}^{11}=\tanh(\mathbf{W}^{11}\mathbf{x}_i+\mathbf{W}^{12}\mathbf{x}_j+\mathbf{b}^1), \\
\mathbf{z}^{12}=\tanh(\mathbf{W}^{11}\mathbf{x}_i+\mathbf{W}^{12}\mathbf{x}_k+\mathbf{b}^1).
\end{aligned}
\end{equation}

Similarly, the outputs of the first layer, $\mathbf{z}^{11}$ and $\mathbf{z}^{12}$, serve as inputs of the second layer. The specific structure of the output of the second hidden layer is expressed as $\mathbf{z}^{21}=\tanh(\mathbf{W}^{2}\mathbf{z}^{11}+\mathbf{b}^2)$ and $\mathbf{z}^{22}=\tanh(\mathbf{W}^{2}\mathbf{z}^{12}+\mathbf{b}^2)$, where $\mathbf{W}^{2}$ represents the weight of the second-layer network, and $\mathbf{b}^2$ denotes the bias. Thus, the final output of the neural network determines the nonlinear function $f$ used to evaluate node similarity in the objective function, which can be expressed as
\begin{equation}
\begin{aligned}
f\left(\mathbf{x}_i, \mathbf{x}_j\right)=\tanh \left(\mathbf{w}^T \mathbf{z}^{21}+b\right),
\end{aligned}
\end{equation}
and
\begin{equation}
\begin{aligned}
f\left(\mathbf{x}_i, \mathbf{x}_k\right)=\tanh \left(\mathbf{w}^T \mathbf{z}^{22}+b\right),
\end{aligned}
\end{equation}
where the elements vector of $\mathbf{w}$ are the weights and the scalar $b$ denotes the bias. The parameter set $\phi$ in the objective function is given by $\phi=\left\{\mathbf{W}^{11}, \mathbf{W}^{12}, \mathbf{W}^2, \mathbf{w}, \mathbf{b}^1, \mathbf{b}^2, b\right\}$, and $H$ is given by $H(\phi)=\left\|\mathbf{W}^{11}\right\|_{F}^{2}+\left\|\mathbf{W}^{12}\right\|_{F}^{2}+\left\|\mathbf{W}^{2}\right\|_{2}^{2} +\|\mathbf{w}\|_{2}^{2}+\left\|\mathbf{b}^{1}\right\|_{2}^{2}+\left\|\mathbf{b}^{2}\right\|_{2}^{2}+b^{2}$.

In the SiNE framework, backpropagation is employed to optimize the deep learning network. This process entails updating network parameters by backpropagating ``errors'', facilitating a more efficient computation of gradients. The key to optimizing the objective function lies in obtaining gradients with respect to the parameters $\mathbf{X}$, $\mathbf{x}_0$, and $\phi$ for $\max(0, f(\mathbf{x}_i, \mathbf{x}_k)+\epsilon-f(\mathbf{x}_i, \mathbf{x}_j))$ and $\max(0, f(\mathbf{x}_i, \mathbf{x}_0)+\epsilon_0-f(\mathbf{x}_i, \mathbf{x}_j))$. Based on the mini-batch stochastic gradient descent algorithm, the training data is divided into small batches during each training iteration. Subsequently, the gradients for the current batch are computed using the backpropagation method. These gradients are then backward propagated from the output layer to the input layer, elucidating the influence of each parameter on the overall network output ``errors''.

\subsection{Node clustering}
After obtaining the embedding vector of each node using SiNE, we further use the K-means algorithm to partition the nodes in the network into $k$ clusters, where $k$ is a tunable parameter. We illustrate the details of using K-means as follows:
\begin{itemize}
\item \textbf{Initialization}: We randomly select $k$ nodes from the signed network, and each of them serves as the central node for one of the $k$ clusters.

\item \textbf{Assignment}: For every node left in the network, we determine the Euclidean distance from it to the cluster centers by utilizing their embedding vectors. We then assign each node to the cluster with the nearest distance and guarantee that each cluster consists of nodes that are most akin to its centroid.
\item \textbf{Update Centroids}: The average of the embedding vectors of the nodes is computed for each cluster, and this average is then designated as the new cluster center.
\item \textbf{Iteration}: The assignment and update centroids steps are iterated until either the cluster centers stabilize or the specified number of iterations is reached.
\end{itemize}

Because each iteration involves a relatively low computational burden, the K-means algorithm runs quickly. By setting the number of clusters ($k$), it promptly aids in selecting nodes and improving the efficiency of the algorithm proposed in this paper.

\subsection{Node Elimination}
Empirical evidence indicates that most nodes are affiliated with a single cluster, while only a minority are assigned to various other distinct clusters. Furthermore, previous studies have indicated that the elimination of nodes within a cluster or community can improve the efficiency of network dismantling~\cite{wandelt2020community,musciotto2023exploring}. Hence, we utilize the largest cluster as the central part for decomposition. More precisely, at each stage of the attack process, we will pinpoint the largest cluster in the network and remove the node with the highest degree in that cluster.
 
\section{Baselines}
\label{3}
To demonstrate the enhanced effectiveness of ESND in network dismantling, we have selected $12$ classic centrality metrics as benchmarks. These metrics encompass those that are agnostic to the sign of the network, such as Degree, Betweenness, K-shell, and Closeness, as well as those that take into account the edge signs, such as P-DEG, N-DEG, Net-DEG, Ratio-DEG, PN, TE, and SPR. The basic explanations of these centrality metrics are provided below.

\begin{itemize}
\item \textbf{Degree}: Degree quantifies the number of direct neighbors of a node when we ignore the sign of the edges, and nodes with higher degrees are generally considered more important. The node degree centrality of node $v_i$ is $\frac {k_i}{N-1}$,
where $N$ represents the number of nodes and $N-1$ signifies the maximum possible degree value for a node, and $k_i$ is the degree of the node $v_i$ indicating the number of its neighbors.

\item \textbf{Betweenness}: It assesses the role of a node in the shortest paths between other nodes. The betweenness centrality of node $v_i$ is $\sum_{i \neq s, j \neq t, s \neq t} \frac{g_{s t}^{i}}{g_{s t}}$, where $g_{s t}$ represents the total number of shortest paths from node $v_s$ to $v_t$, and $g_{s t}^{i}$ denotes the number of these shortest paths among the $g_{s t}$ that pass through $v_i$.

\item \textbf{K-shell}: K-shell centrality categorizes nodes based on their degrees to evaluate their importance in a network. Assuming there are no isolated nodes in the network, we eliminate nodes with one connection until no more such nodes remain and assign them to the $1$-shell. Similarly, we recursively eliminate nodes with degree of $2$ to form the $2$-shell. This process concludes when all nodes have been allocated to one of the shells.

\item \textbf{Closeness}: This centrality functions as a global indicator delineating the node's position in the network, and it quantifies the average distance between a node and the remaining nodes. The closeness centrality of node $v_i$ is $\frac{N-1}{\sum_{j\neq i}d_{ij}}$, where $d_{ij}$ is the length of the shortest path between node $v_i$ to node $v_j$. A higher closeness value indicates that $v_i$ is closer to the other nodes in a network. 

\item \textbf{Positive degree (P-DEG)}: P-DEG counts the number of positive edges linked to a node, which is referred to as the positive degree. Thus, the P-DEG centrality value of node $v_i$ is given by its number of positive edges ${k_{i}^{+}}$.

\item \textbf{Negative degree (N-DEG)}: N-DEG quantifies the number of negative edges associated with each node, denoted as the negative degree. The N-DEG centrality of node $v_i$ can be presented by its number of negative edges ${k_{i}^{-}}$.

\item \textbf{Net degree (Net-DEG)}: This metric represents the difference between the number of positive edges and negative edges that a node has. For node $v_i$, the Net-DEG value is presented as ${k_{i}^{+}}-{k_{i}^{-}}$.

\item \textbf{Ratio degree (Ratio-DEG)}: It represents the proportion of positive edges that a node $v_i$ has among its total number of edges in the network, which reads $\frac{k_{i}^{+}} {{k_{i}^{+}}+{k_{i}^{-}}}$. 

\item \textbf{PN centrality}~\cite{everett2014networks}: Everett and Borgatti argue that nodes with more positive connections are more significant, while nodes with more negative connections are less important. Thus, they propose the PN index to evaluate node importance in signed networks, calculated using the following formula
\begin{equation}
\begin{aligned}
PN=\left(I-{\frac{1}{2N-2}}A\right)^{-1} \textbf{1},
\end{aligned}
\end{equation}
where $N$ represents the number of nodes in the network, $I$ is the $N$-order identity matrix, $A = A^{+} - 2A^{-}$, and $A^{+}$ (or $A^{-}$) represents the adjacency matrix containing only positive (or negative edges). \textbf{1} denotes an $N$-dimension vector with all elements equal to 1.

\item \textbf{TE}~\cite{liu2020simple}: This index calculates the centrality of a target node considering the total effect (TE) of all other nodes to it in the network. The higher the value of TE, the more important the node. For an undirected signed network, if there is an edge between $v_i$ and $v_j$, the effect of $v_i$ to $v_j$ is defined as $E_{ij,1^S} = S \times \frac{1}{D_{j}}$, where $S$ is the sign ($+1$ or $-1$) of the edge $e_{ij}$, and $D_{j}$ is the degree of $v_j$. We construct two matrices $CE_{n^+}=\{CE_{ij,n^+}\}_{N\times N}=\{\sum_{l=1}^{n}E_{ij,l^+}\}_{N\times N}$ and $CE_{n^-}=\{CE_{ij,n^-}\}_{N\times N}=\{\sum_{l=1}^{n}E_{ij,l^-}\}_{N\times N}$ to represent the sum of the positive and negative effects of $v_i$ to $v_j$ up to $n$ steps, respectively. Therefore, $TE_{ij,n}=CE_{ij,n^+} + \left| CE_{ij,n^-} \right|$ indicates the sum of effects from $v_i$ to $v_j$, and the TE value of $v_i$ is further given by 
\begin{equation}
\begin{aligned}
TE_{i,n}=\sum_{j=1}^{N}TE_{ij,n}.
\end{aligned}
\end{equation}

Here, we set $n=2$, meaning that we only calculate the effect of a node on its neighbors in two hops.

\item \textbf{Signed-PageRank (SPR)}~\cite{yin2019signed}: SPR is a PageRank algorithm adapted for signed networks, which updates the SPR value for each node in each iteration by aggregating the weights and sign information. The formula for updating the SPR value of $v_i$ at iteration $t+1$ is

\begin{equation}
\begin{aligned}
SPR_{i,t+1}=\sum_{v_j\in D_{i}^{out}}(SPR_{i,t}-SPR_{j,t}){y_{i,j}}+\frac{1-d}{N},
\end{aligned}
\end{equation}

where $D_{i}^{out}$ is the set of out-neighbors of $v_i$. $Y=\{y_{i,j}\}_{N\times N}=dH$ is the Signed-PageRank adjacency matrix with damping coefficient $d$, where $H$ represents the Hadamard product of the normalized weight matrix $W$ and the label matrix $L$. In our work, the weights of all edges are equal to $1$, thus in matrix $W=\{w_{ij} \}_{N\times N}$, ${w_{ij}}=\frac{1}{D_i}$ if $v_i$ and $v_j$ have a connection. In matrix $L=\{l_{ij}\}_{N\times N}$, $l_{ij} = 1$ if there is a positive connection between $v_i$ and $v_j$, and $l_{ij} = -1$ signifies a negative connection between them. Unlike the PageRank algorithm, the iteration of the Signed-PageRank algorithm continues until the ranking of nodes based on SPR values remains unchanged. Here, we consider the final rank of the nodes as their importance.

\item \textbf{Signed Eigenvector (SE)}~\cite{bonacich2004calculating}: SE is an extension of the eigenvector of signed networks. The main idea is that a node with more positive edges to the important nodes is more important, and vice versa for nodes with more negative edges to the important nodes. Given the label matrix $L_{N\times N}$ of an undirected and unweighted signed network, we can swap the rows and columns of $L$ to obtain a matrix
\begin{equation}
\begin{aligned}
A=\left(\begin{matrix}L^{+} & L^{-}\\
L^{-} & L^{+}
\end{matrix}
\right),
\end{aligned}
\end{equation}
where $L^+_{n_1 \times n_1}$ is an adjacency matrix only containing positive edges, $L^{-}_{n_2 \times n_2}$ denotes a adjacency matrix with negative edges, and $n_1 + n_2 = N$. Let $B=DAD$, where $D$ is a diagonal matrix, whose first $n_1$ diagonal elements are $1$ and the remaining $n_2$ elements are equal to $-1$. In particular, $B$ has positive eigenvalues $\lambda$ because it contains only non-negative elements and corresponding eigenvector $x$. Since $Bx=DADx=\lambda x$, we have $ADx = \lambda D^{-1}x =\lambda Dx$. Therefore, the signed eigenvalue centrality of each node can be represented by the eigenvector $Dx$ when $Dx$ is in a steady state.



\end{itemize}

\section{Experiments}
\label{4}
We apply the proposed ESND to dismantle six distinct real signed networks and three different signed network null models, and compare the results of ESND with those of the baselines on these null models to assess the stability of ESND. Additionally, we compute Kendall correlation coefficients for target attack node sequences generated by various decomposition strategies to analyze differences in node selection. Finally, we test how the ratio of negative edges would affect the robustness of a signed network through artificial network models.

\subsection{Datasets}
We select six real-world datasets that can be constructed as signed networks to evaluate the performance of our method. Specifically, \textbf{Bitcoinalpha} and \textbf{Bitcoinotc} are data sourced from SNAP \footnote{https://snap.stanford.edu/data/\label{fn:1}}, illustrating the trust networks between users participating in Bitcoin transactions. Due to transactional anonymity in Bitcoin transactions, users provide positive and negative ratings to signify trust (positive) or distrust (negative) relationships. \textbf{WikiVote} represents the voting network to select Wikipedia administrators\footnote{https://doi.org/10.6084/m9.figshare.12152628\label{fn:2}}. The eligibility of the users for administration is determined through voting, with the edges denoting voting interactions, i.e., positive signs indicate support while negative signs indicate opposition. \textbf{Slashdot} is a notable technology news site where users comment and share technology-related information\footnote{https://www.aminer.cn/data-sna\label{fn:3}}. Positive and negative signs in the dataset denote friendly or adversarial relationships between users. \textbf{Reddit} captures connections between users in diverse sub-communities, reflecting positive or negative sentiment in shared content across online accounts\footref{fn:2}. \textbf{Epinions} constitutes a trust network among users on a product review website, with positive and negative signs indicating trust or distrust relationships between user connections\footref{fn:3}. We show the topological information of these signed networks in Table~\ref{table1}, including the number of nodes ($N$), the number of edges ($M$), the number of positive edges ($E^{+}$), the number of negative edges ($E^{-}$), the average degree of nodes ($\left\langle k \right\rangle$) and the clustering coefficient ($C$). The table shows that all the signed networks have more positive edges than negative ones.

\begin{table}[!hbp]
\renewcommand{\arraystretch}{1.2}
\centering
\caption{Topological information of the signed networks, in which $N$ denotes the number of nodes; $M$ represents the number of edges; $\left| E^{+} \right|$ and $\left| E^{-} \right|$ indicate the number of positive and negative edges, respectively. The values in parentheses represent the proportions of positive and negative edges in the network; $\left\langle {k} \right\rangle$ denotes the average degree, and $C$ signifies the average clustering coefficient.}
\resizebox{1.0\linewidth}{!}{
\begin{tabular}{ccccccc}
\toprule
\multicolumn{1}{l}{}  & \textbf{$N$} & \textbf{$M$} & \textbf{$\left| E^{+} \right|$} & \textbf{$\left| E^{-} \right|$} & {$\left\langle {k} \right\rangle$} & \textbf{$C$} \\ \hline
Bitcoinalpha & 3783           & 14124          & 12759(90\%)             & 1365(10\%)              & 7.47                                & 0.177                 \\
Bitcoinotc   & 5881           & 21492          & 18250(85\%)             & 3242(15\%)              & 7.31                                & 0.178                 \\
WikiVote    & 7118           & 100751         & 78658(78\%)             & 22093(22\%)             & 28.3                                & 0.141                 \\
Slashdot      & 13182          & 34260         & 28884(84.3\%)           & 5376(15.7\%)             & 5.19                               & 0.149                 \\
Reddit     & 18282          & 107301          & 99084(92.3\%)           & 8217(7.7\%)            & 11.74                                & 0.374                  \\
Epinions     & 25148          & 99880          & 69185(69.2\%)           & 30695(30.7\%)           & 7.94                                & 0.073                 \\ \bottomrule
\end{tabular}
}
\label{table1}
\end{table}

\subsection{Performance Evaluation Metric}
Network dismantling methods aim to produce an optimal node sequence to remove that could disrupt the network as much as possible. We use the robustness metric $R$ to assess the performance of ESND, as well as the baselines\cite{schneider2011mitigation,wandelt2018comparative}
\begin{equation}
\begin{aligned}
R=\frac{1}{N} \sum_{Q=1}^{N}s(Q),
\end{aligned}
\end{equation}
where $N$ is the size of network, $s(Q)$ represents the fraction of nodes in the largest connected components after the removal of $Q=qN$ nodes, and $1/N$ is a standardized operation for comparing the robustness of networks with different sizes. To compute $R$, a node rank is necessary; therefore, various dismantling methods are proposed to find the minimum $R$ in all possible node orders. A lower value of $R$ indicates that the method is more effective in destroying the network.

\subsection{Parameter Analysis}

\begin{figure*}[!htb]
    \centering
        \includegraphics[width=1.0\textwidth]{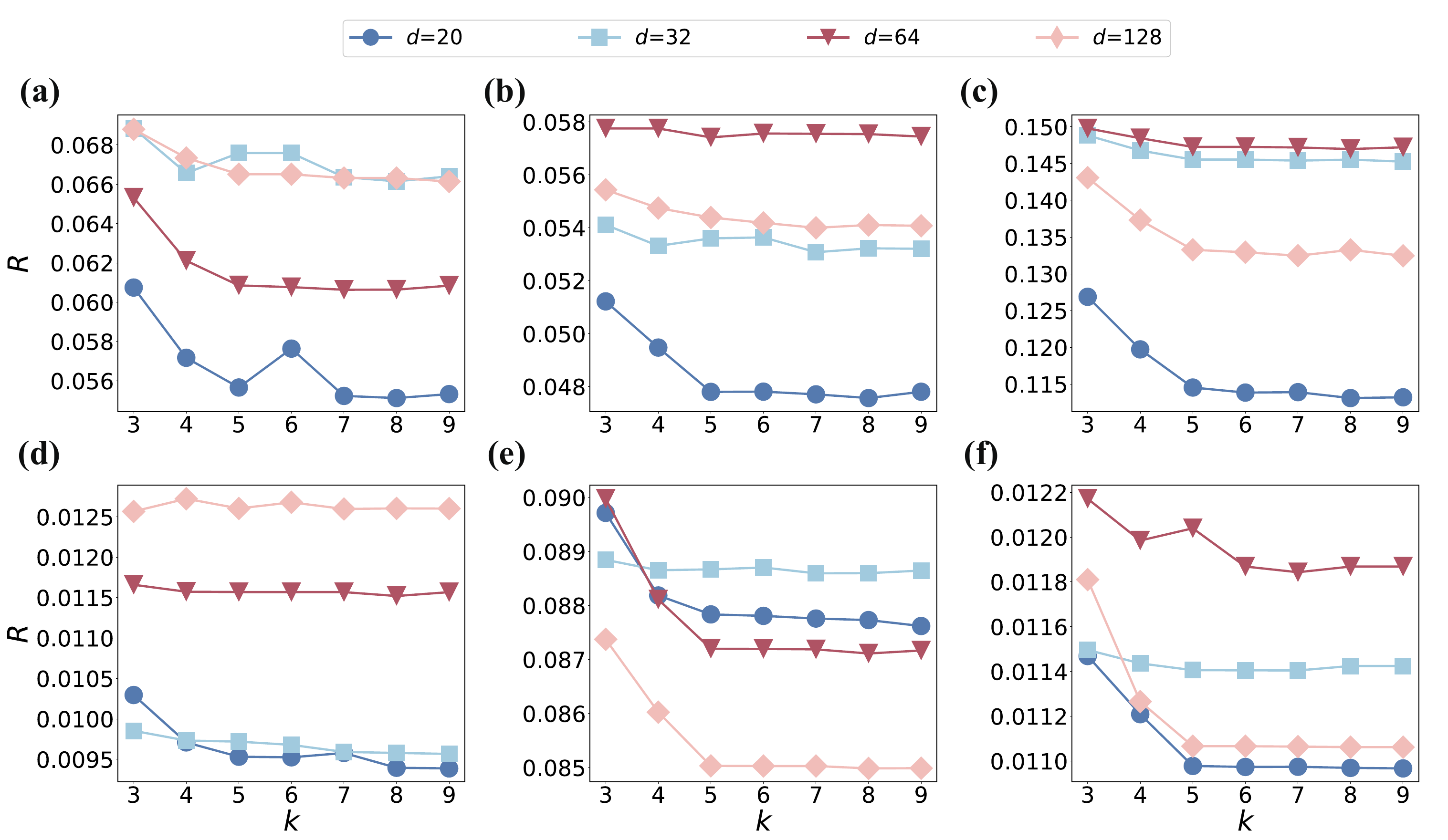}
    \caption{Performance of ESND under different parameter settings. The x-axis indicates the number of clusters $k$ and the y-axis shows the performance of network dismantling. Different curves show the use of different values of dimension $d$ in the embedding procedure. We show results for: (a)Bitcoinalpha; (b)Bitcoinotc; (c)WikiVote; (d)Slashdot; (e)Reddit; (f)Epinions.}
    \label{fig2}
\end{figure*}

To optimize the effectiveness of dismantling the network of the proposed method, we perform a thorough analysis of various parameters. Specifically, we focus on two key parameters, i.e., the embedding dimension size $d$ and the number of clusters $k$, and keep the other parameters unchanged (we set hidden layers $L=2$, learning rate $\lambda=0.0001$, and similarity parameters $\epsilon$ and $\epsilon_0$ set to 1 and 0.5, respectively. Note that these parameters are unchanged in the following experiments). We systematically compare the $R$ values for each dataset across different values of $d$ and $k$, the results are given in Figure \ref{fig2}. We observe that when $k$ is unchanged, the smallest $R$ is given by $d=20$ in most networks, except \textbf{Reddit} where $d=128$ achieves the best performance. Meanwhile, as $k$ increases, the value of $R$ decreases and reaches its minimum at $k=8$ in all networks. Therefore, in the following experiments, we set $k=8$ for the six networks, $d=128$ for \textbf{Reddit}, and $d=20$ for the remaining networks.

\subsection{Experimental Results}
\subsubsection{Results on Real Signed Networks}
\begin{figure*}[!htb]
    \centering
        \includegraphics[width=1.0\textwidth]{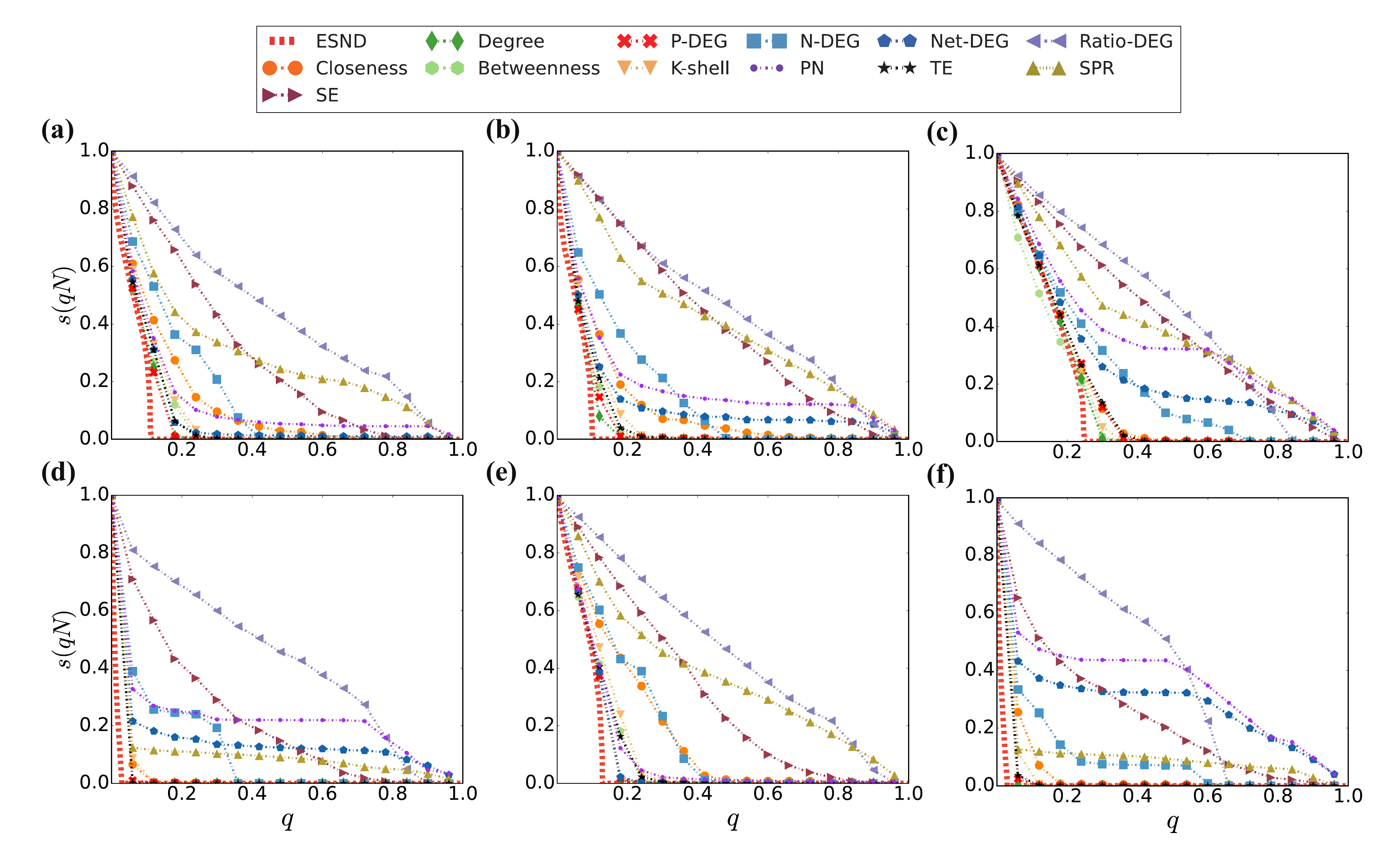}
    \caption{Comparison of ESND with baselines on signed networks: (a)Bitcoinalpha; (b)Bitcoinotc; (c)WikiVote; (d)Slashdot; (e)Reddit; (f)Epinions. X-axis shows the fraction of nodes removed and y-axis means the ratio of nodes in the giant component after node removal.}
    \label{fig3}
\end{figure*}

We compare the performance of the ESND with the selected baselines on the six signed networks, where the results are given in Figure~\ref{fig3} and Table~\ref{table2}. In Figure~\ref{fig3}, the horizontal axis ($q$) represents the proportion of nodes removed, while the vertical axis ($S(qN)$) corresponds to the fraction of nodes in the GCC after removing $q$ fraction of nodes. For a fixed value of $q$, the smaller value of $S(qN)$ indicates that the dismantling method is more effective in dismantling the corresponding signed network than other methods. The values in Table~\ref{table2} reveal the area under each curve (AUC) in Figure~\ref{fig3}, with a smaller value indicating the better performance of the corresponding dismantling method. The experimental results show that the robustness of these real signed networks is notably different, with some of them demonstrating fast network collapse with only a small fraction of nodes being removed, such as \textbf{Slashdot} and \textbf{Epinions}, while the remaining ones are more robust. For example, \textbf{WikiVote} and \textbf{Reddit} networks necessitate approximately 40\% removal to attain complete decomposition for most of the dismantling methods. In addition, ESND outperforms all baseline methods in dismantling most signed networks, particularly when we remove a large fraction of nodes. In dismantling an unsigned network, normally the betweenness can outperform the other methods in most cases\cite{wandelt2018comparative}. However, it performs second best in most cases in dismantling signed networks, which reveals that considering the topology deduced by the signs is important in dismantling a signed network. Moreover, various centrality methods exhibit varying performances across different datasets, including Closeness, K-shell, PN, TE, SPR, and SE. The efficacy of these methods is closely related to the specific structures of the networks. In contrast, ESND consistently achieves optimal network dismantling results across diverse datasets, showing its stability and effectiveness.

\begin{table*}[!t]
\renewcommand{\arraystretch}{1.2}
\centering
\caption{Area under each curve (AUC) of each curve in Figure~\ref{fig3}. The best performance is highlighted in bold.}
\resizebox{1.0\textwidth}{!}{
\begin{tabular}{cccccccccccccc}
\toprule
             & ESND   & Degree & P-DEG  & N-DEG  & Net-DEG & Ratio-DEG & Closeness & Betweenness & K-shell & PN     & TE     & SPR    & \textbf{SE}     \\ \hline
Bitcoinalpha & \textbf{0.0596} & 0.0704 & 0.0699 & 0.1603 & 0.0907  & 0.4356    & 0.1305    & 0.0782      & 0.0949  & 0.1327 & 0.0823 & 0.2955 & 0.2972 \\
Bitcoinotc   & \textbf{0.0499} & 0.0579 & 0.0611 & 0.1599 & 0.1326  & 0.4538    & 0.1133    & 0.0634      & 0.0789  & 0.1889 & 0.0692 & 0.404  & 0.4021 \\
WikiVote     & 0.1455 & 0.1521 & 0.1644 & 0.2315 & 0.2686  & 0.4562    & 0.1675    & \textbf{0.1363}      & 0.1596  & 0.3610 & 0.1649 & 0.4067 & 0.4274 \\
Slashdot     & \textbf{0.0106} & 0.0128 & 0.0161 & 0.0975 & 0.1357  & 0.4246    & 0.022     & 0.0119      & 0.0289  & 0.2087 & 0.0137 & 0.0958 & 0.2186 \\
Reddit       & \textbf{0.0802} & 0.0921 & 0.0915 & 0.1778 & 0.0935  & 0.4616    & 0.1774    & 0.1033      & 0.1152  & 0.1097 & 0.1027 & 0.3748 & 0.3173 \\
Epinions     & \textbf{0.0097} & 0.0123 & 0.0161 & 0.0877 & 0.2774  & 0.4003    & 0.0457    & 0.0108      & 0.0256  & 0.3428 & 0.0147 & 0.0989 & 0.2363 \\ \bottomrule
\end{tabular}
}
\label{table2}
\end{table*}

\begin{figure*}[!htb]
    \centering
        \includegraphics[width=1.0\textwidth]{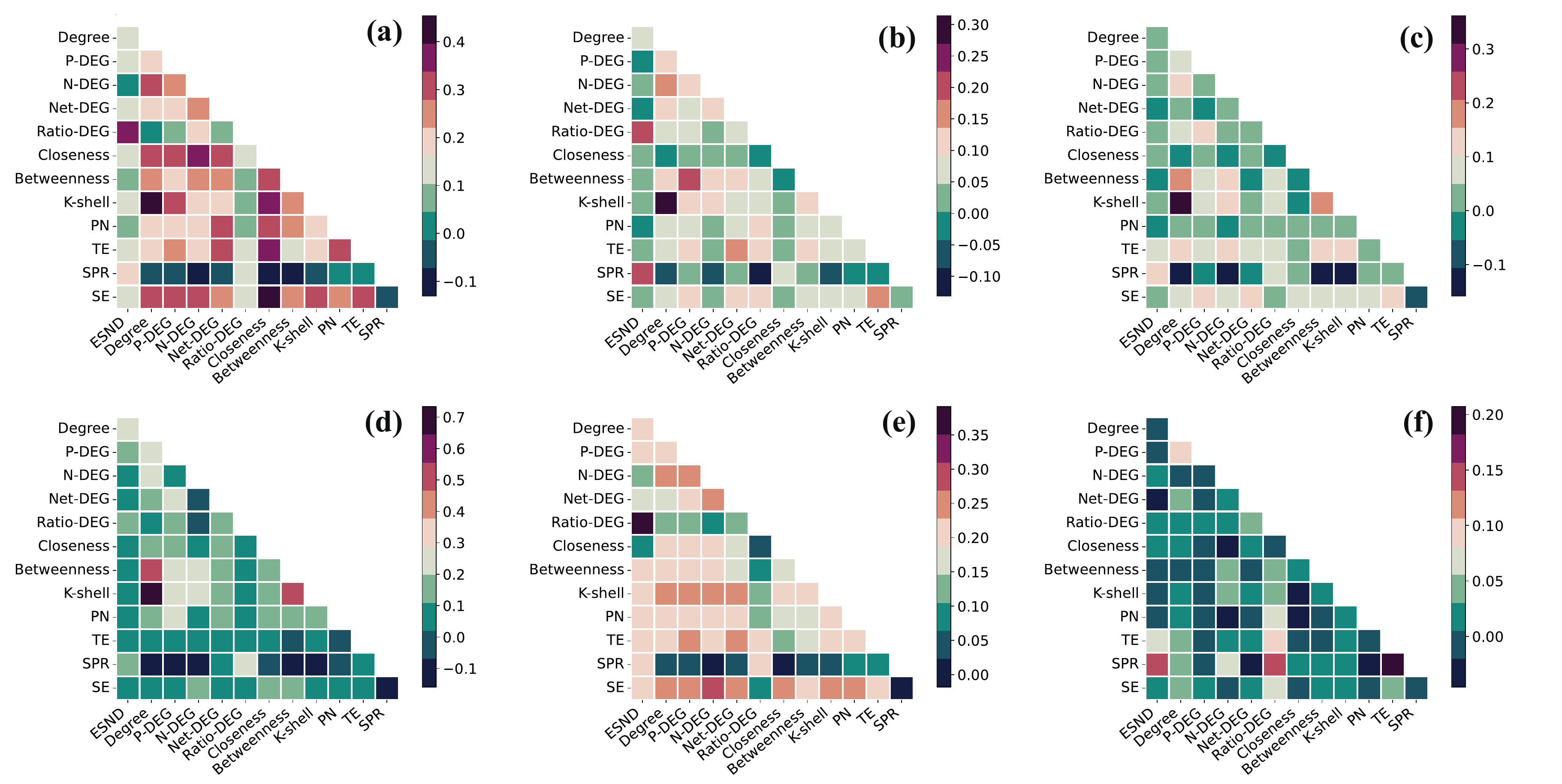}
    \caption{Analysis of differences between removal node sequences generated by different methods. Each square represents the Kendall correlation coefficient between the removed node sequences generated by the corresponding pair of methods. We show the results for the following signed networks: (a) Bitcoinalpha; (b) Bitcoinotc; (c) WikiVote; (d) Slashdot; (e) Reddit; (f) Epinions.}
    \label{fig4}
\end{figure*}

Various methods demonstrate diverse efficacy in network dismantling due to disparities in the strategies employed for node selection during each iteration. To scrutinize the dissimilarities in the node removal sequences generated by these methods, we conduct a correlation analysis. For each method, we first obtain the node removal sequence, i.e., different methods may result in different orders of node removal. Then we calculate the Kendall correlation coefficient between the node sequences obtained by a pair of dismantling methods. The Kendall correlation coefficients between each pair of methods are given in Figure~\ref{fig4}.
In particular, the Kendall correlation coefficients between the proposed ESND and the baselines are generally low, indicating a significant deviation in the node removal strategy of the ESND from these baseline methods. Additionally, Degree, P-DEG, N-DEG, Net-DEG, and Ratio-DEG, despite relying on distinct dismantling strategies derived from node degree, yield node sequences with relatively low correlation.  

\subsubsection{Results on Null Models of the Signed Network}
\begin{figure}[!t]
    \centering
        \includegraphics[width=1.0\linewidth]{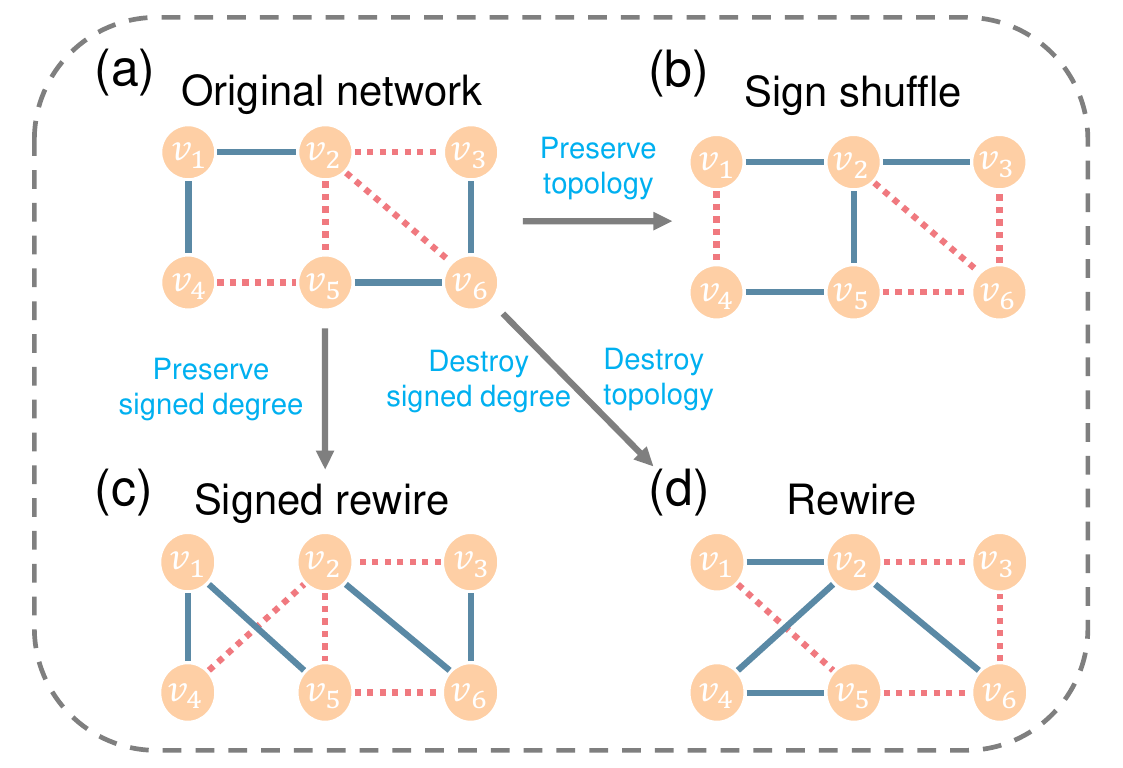}
    \caption{Toy examples of null models of a signed network. Solid lines represent positive edges, and dotted lines represent negative edges.}
    \label{fig5}
\end{figure}

To delve deeper into the potential influence of factors such as network topology and signs on ESND and their consequent impact on variations in network dismantling outcomes, three distinct null models for signed networks\cite{hao2024proper} were constructed in six datasets. We illustrate examples of the null models in Figure~\ref{fig5}, in which they preserve certain properties of the original network. Detailed descriptions of them are given below.

\begin{itemize}
\item \textbf{Sign shuffle:} In this model, the topological structure of the network is preserved by randomly selecting one positive edge and one negative edge and exchanging their signs, but the positive and negative degrees of each node will change. Taking node $v_1$ in Figure~\ref{fig5}a and b as an example, the degree of node $v_1$ is preserved, but the positive degree and negative degree of node $v_1$ change from $\{2,0\}$ to $\{1,1\}$ via the sign shuffle model.

\item \textbf{Signed rewire:} Initially, two subgraphs containing only positive or negative edges are constructed from the original network. Subsequently, the edges are rewired within each subgraph, which could preserve the positive and negative degrees of the nodes.
 The process ends by merging the two rewired subgraphs to establish the null model. In this model, the positive and negative degrees of the nodes remain the same as in the original network, while the network structure is changed. Figure~\ref{fig5}c demonstrates the generation of a signed rewire null model. For example, we disconnect the edges $(v_1, v_2)$, $(v_5, v_6)$ and form new edges $(v_1, v_5)$, $(v_2, v_6)$ but keep the positive and negative degree of each node. 
\item \textbf{Rewire:} The model exchanges edges between nodes while keeping the degree of each node unchanged. In this null model, both the topological structure of the network and the positive and negative degrees of each node undergo alterations. In Figure~\ref{fig5}d, we show that the degree, positive degree, and negative degree of each node are changed through the random rewiring process of the rewire model.

\end{itemize}
\begin{figure*}[!htb]
    \centering
        \includegraphics[width=1.0\linewidth]{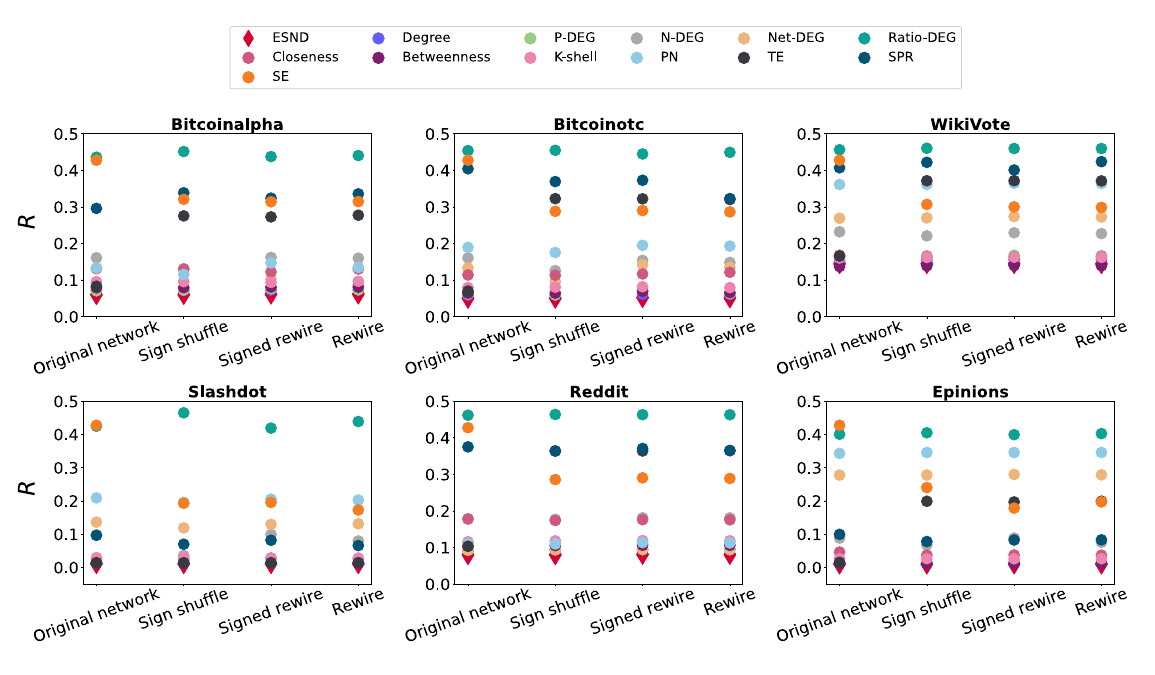}
    \caption{Network dismantling on different signed network null models. We show the results for signed networks: Bitcoinalpha; Bitcoinotc; WikiVote; Slashdot; Reddit and Epinions. The results are the average of 100 realizations.}
    \label{fig6}
\end{figure*}

We perform network dismantling on these null models generated by each of the six signed real-world networks, the specific experimental results are illustrated in Figure~\ref{fig6}. The horizontal axis denotes the original signed network and their corresponding null models, while the vertical axis represents the evaluation metrics $R$ for various network dismantling methods applied to these signed networks and null models. The results show that each of the dismantling methods demonstrates generally consistent performance in network dismantling across both the original network and three null models within the same dataset. This suggests that modifications to the topological properties and sign distribution of the signed network do not significantly affect the efficacy of these methods. More importantly, ESND consistently attains superior dismantling performance compared to these baselines across these varied null models (as shown by the red diamonds in the figures), emphasizing the stability of ESND as an effective method for dismantling networks.

\subsubsection{Impact of the signs on network robustness}
\begin{figure}[!t]
    \centering
        \includegraphics[width=1.0\linewidth]{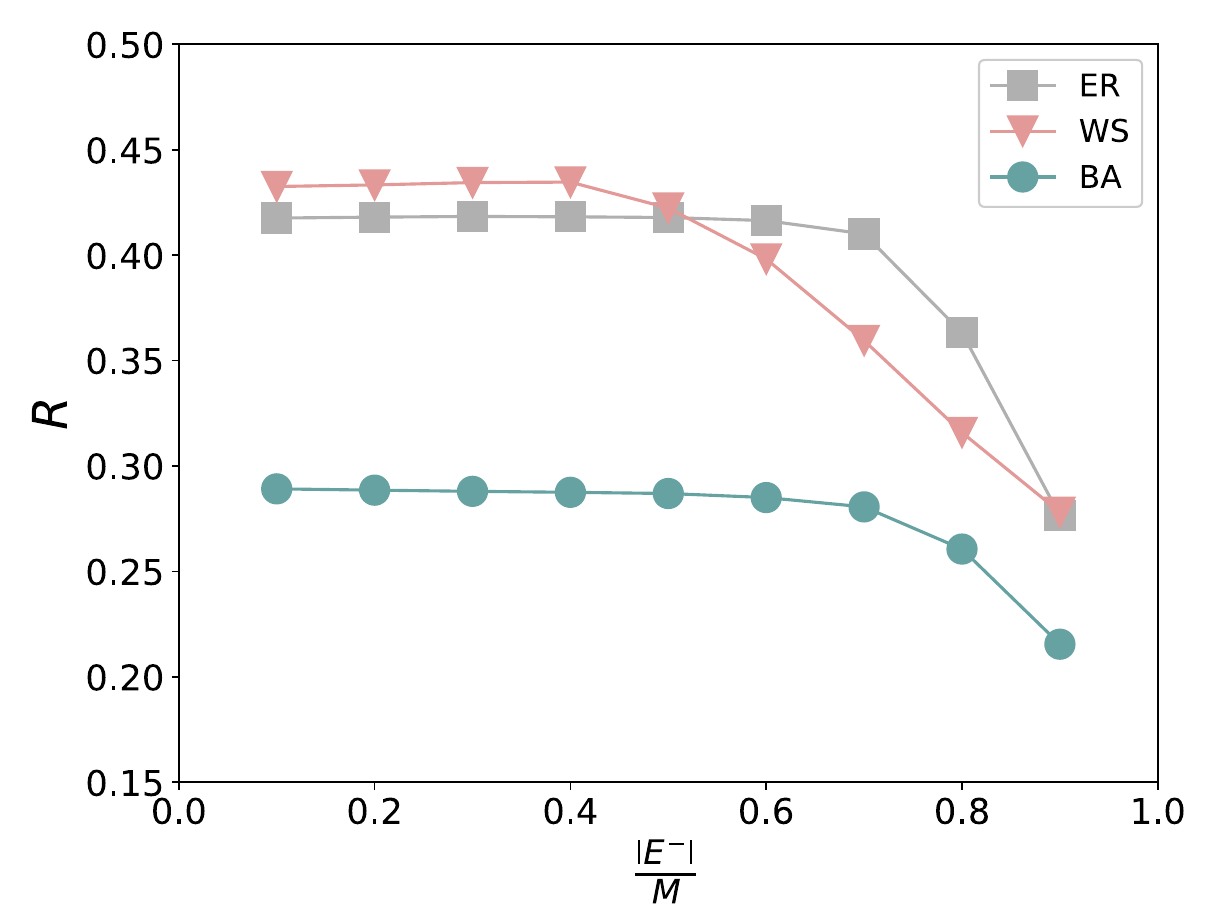}
    \caption{Robustness of synthetic signed network with the change of negative edge ratio. The horizontal axis represents the proportion of negative edges, the vertical axis represents the $R$, and different lines correspond to different synthetic networks, i.e., ER, WS and BA. Each point is averaged over 100 realizations.}
    \label{fig7}
\end{figure}

We further examine the robustness of a signed network by adding different ratios of positive or negative edges. Specifically, we first generate unsigned synthetic networks, i.e., ER, WS, and BA, and then assign different ratios of positive or negative edges in the networks. Finally, we evaluate the robustness of these networks by using ESND. To be consistent and comparable, all synthetic networks contain 1000 nodes with the same average degree of $10$. In the ER network, the probability of randomly connecting edges is set to $p=0.01$. For the WS network, each node is connected to its $k=10$ nearest neighbors, with a rewiring probability of $p=0.01$. In the BA network, the initial number of nodes is $m_{0}=6$, and each new node was connected to $5$ existing nodes. Subsequently, random positive and negative signs were assigned to each edge in each synthetic network, controlling the ratio of negative edges $p_{-}=[0.1,\cdots,0.9]$ to generate signed synthetic networks corresponding to different negative edge ratios. We show the dismantling results in Figure~\ref{fig7}, where each point is the average of $100$ realizations.

In Figure~\ref{fig7}, the x-axis indicates the ratio of negative edges ($\frac{|E^{-}|}{M}$) in each of the networks, and the y-axis shows the $R$ values, revealing the robustness of the corresponding networks. Although the WS network has a higher value of $R$ (indicating more robustness) for a low value of $\frac{|E^{-}|}{M}$ compared to ER and BA, it is easier to disassemble when $\frac{|E^{-}|}{M}>0.4$. Meanwhile, we observe that as the ratio of negative edges increases for a relatively small value of $\frac{|E^{-}|}{M}$ ($\frac{|E^{-}|}{M}<0.4$ for WS, $\frac{|E^{-}|}{M}<0.7$ for ER and BA), the robustness of the networks is relatively stable. However, for a large value of $\frac{|E^{-}|}{M}$, the networks can easily be dismantled, with the value of $R$ decreasing with increasing $\frac{|E^{-}|}{M}$. This suggests that increasing the proportion of positive edges in the network contributes to enhancing its robustness. This observation aligns with real-world scenarios. In a social network where negative edges dominate, signifying antagonistic relationships between individuals, the network is naturally more vulnerable. In general, ER and BA networks are more resilient than the WS network with increasing $\frac{|E^{-}|}{M}$. 

\section{Conclusion}
\label{5}
In this study, we propose an embedding-based network dismantling framework, namely ESND, to address the signed network dismantling problem. The algorithm mainly iteratively processes the following four steps: it first detects the giant connected component (GCC) within the network and then utilizes the signed network embedding algorithm (SiNE) to generate embedding vectors for each of the nodes. Then, it partitions the GCC into different groups via the K-means algorithm based on the node embedding vectors. Subsequently, the node with the highest degree in the largest cluster is removed from the network. The above process is repeated until the fraction of removed nodes, indicated as $q$, reaches a predetermined threshold $q_{r}$. Comprehensive experiments on various real signed networks and their corresponding null models demonstrate that ESND surpasses other baseline methods, thus confirming its efficacy and stability. Additionally, correlation analysis of the removed node sequences reveals why ESND can better dismantle a signed network than other baseline methods. Moreover, experiments with testing how the ratio of negative edges in a network could affect the robustness of a signed network show that networks with more negative edges are easier to dismantle.

Certain aspects of this study warrant further attention. Future work could explore the following aspects: a) in this work, we treat signed networks as undirected networks. However, most of the signed networks in the real world contain directionality. Therefore, how to efficiently dismantle a directed signed network remains a topic worth investigating. b) We confine ourselves to using unsigned network connectivity, i.e., the fraction of nodes in the giant component, to evaluate the performance of a dismantling method. Future work could also propose performance evaluation methods regarding the sign nature of a network. c) Most of the existing research on network dismantling focuses on removing critical nodes. Devising methods that could identify critical edges in signed networks to achieve rapid network decomposition is also an interesting direction that deserves further exploration.

\section*{Acknowledgment}
This work is supported by the National Natural Science Foundation of China (Grant Nos. 62173095, U23A20331), Natural Science Foundation of Zhejiang Province (Grant Nos. LQ22F030008), Scientific Research Foundation for Scholars of HZNU (2021QDL030), and the Natural Science Foundation of Shanghai (Grant No. 21ZR1404700).

\bibliographystyle{IEEEtran}
\bibliography{ref}
\end{document}